\baselineskip=18pt
\font\ninerm=cmr9
\font\ninemi=cmmi9
\font\ninebx=cmbx9

\def\hs{\hskip}

\def\b#1{{\bf #1}}
\def\r#1{{\rm #1}}
\def\h{\hat}
\def\s#1{{^{*}#1}}
\def\d#1{{\delta #1}}

\line{November 1997\hfil ENSK-ph/97-03}
\bigskip\bigskip\bigskip\bigskip\bigskip
\centerline{\bf A POTENTIAL INFRARED PROBLEM WITH THE DAMPING RATES} 
\centerline{\bf  FOR GLUONS WITH SOFT MOMENTUM IN HOT QCD}
\bigskip\bigskip
\centerline{A. Abada, O. Azi and K. Benchallal}
\medskip
\centerline{\it  D\'epartement de Physique, \'Ecole Normale
                Sup\'erieure,}
\centerline{\it BP 92, Vieux Kouba 16050, Alger, Algeria}
\medskip
\centerline{\tt enskppps@ist.cerist.dz}
\bigskip\bigskip\bigskip
\centerline{\ninebx Abstract}
\bigskip
\hbox{\hs1truein\vbox{\hsize=4.5truein\baselineskip=12pt\ninerm
We calculate the damping rate $\ninemi\gamma_l$ for longitudinal
gluons with zero momentum in finite high temperature QCD and
show that some of its contributing terms are infrared divergent.
This is in contrast with the expectation that this damping rate
is to be equal to the corresponding one $\ninemi\gamma_t$
for transverse gluons which is known to be finite.
Our calculation was motivated by the fact that similar
divergent terms occur when we calculated in a previous work
$\gamma_t$ to order $\ninemi p^2$, {\ninebx p} being
the momentum of the gluon. After we present our results,
we briefly discuss them.}}
\vfil\eject

In a previous work [1], we undertook the calculation of the damping rate
$\gamma_t(p)$ for transverse gluons with soft momentum $\b p$ in
finite-high-temperature QCD. Knowing that $\gamma_t$ at zero momentum is
finite [2], our main aim in that work was to argue that for small enough
$p=|\b p|$, the expansion of $\gamma_t(p)$ in powers of $(p/m_g)^2$ is valid,
$m_g$ being the inverse gluonic correlation length. We suggested that,
when contrasted with a previously recent calculation [3]
based on letting $p$ come from the hard limit (set by the temperature $T$)
down towards the interior of the soft region (set by $m_g\sim gT$ where $g$
is the QCD coupling constant), a calculation that produced a behavior
of the form $\ln(1/g)$ for $\gamma_t(p)$, such a result could possibly be
indicative of a kind of transition around some `critical' scale.
We therefore expanded $\gamma_t(p)$ in the following manner:
$$
 \gamma_t(p)={g^2N_cT\over 24\pi}\
 \Big[\ a_{t0}+a_{t1}\Big({p\over m_g}\Big)^2+\dots\Big]\ ,\eqno(1)
$$
and gave an analytic expression for $a_{t1}$ (the first coefficient
$a_{t0}$ was already calcualted in [2]). $N_c$ is the number of colors.

The natural step forward was to calculate numerically $a_{t1}$. But quite
interestingly, when we started manipulating the expression of $a_{t1}$
we obtained, we realized that some of its contributing terms
are infrared divergent (in a sense that will become clear later in the text).
Thus, on the one hand, the quick conclusion is that $a_{t1}$ may be infinite
and hence, the expansion in (1) is not valid after all. On the other
hand, such a potential infrared divergence seems in contradiction with
the physical expectation that the quark-gluon plasma is stable for,
at least, very small gluonic momenta.

In order to look into the matter more closely, we undertake in this work the
derivation of $\gamma_l(0)$, the damping rate for longitudinal gluons at
zero momentum. Our motivation is that, first, this damping
rate is to be equal to $\gamma_t(0)$, simply because at zero momentum, there
is no difference between transverse and longitudinal directions.
Second, the interesting point in this calculation is that
in order to get $\gamma_l$ at zero $p$, it is already necessary
to expand it to order $(p/m_g)^2$. After the calculation is performed,
we indeed recover the finite analytic expression that is equal to
$\gamma_t(0)$, but, surprisingly enough, together with additional terms quite
similar in form to the ones we obtained in the case of $\gamma_t(p)$,
terms when taken individually present an infrared divergence.
To the best of our knowledge, in this context, such infrared-divergent
terms have not been discussed previously in the literature.
Before furthering this discussion, we first present in the sequel
our results.

In the imaginary-time formalism, the eucledian momentum of the gluon
is $P^\mu = (p_0 , \b p)$ such that $P^2 = (p_0 )^2 + p^2 $
with $\b p =p\; \b{\h{p}}$ and $p_0 = 2\pi nT$ where $n$ is an integer.
Real-time amplitudes are obtained via the analytic continuation
$p_0 = -i\omega + 0^+ $ where $\omega$ is the energy of the gluon. 
A momentum is said to be soft if both $\omega$ and $p$  are of order $gT$;
it is said to be hard if one is or both are of order $T$.
In the Coulomb gauge, the complete inverse gluon propagator is:
$$
 D_{\mu\nu}^{-1}(P) = P^2 \delta^{\mu\nu} - P^{\mu}P^{\nu}-\Pi^{\mu\nu}(P)
 + {1\over \xi_{C}} \delta^{\mu i}\delta^{\nu j} p^{i}p^{j}\ , \eqno(2)
$$
where $\Pi^{\mu\nu}(P)$ is the gluon self-energy and the last term  
is due to Coulomb-gauge fixing. We work in the strict 
Coulomb gauge $\xi_{C}=0$. The gluon self-energy can be decomposed into:
$$
 \Pi^{\mu\nu}(P)=\d{\Pi}^{\mu\nu}(P) + \s{\Pi}^{\mu\nu}(P)\ , \eqno(3)
$$
where $\d{\Pi}$ is the hard-thermal-loop and 
$\s{\Pi}$ is the effective self-energy [4]. $P$ being soft, 
the hard thermal loop is of the same order of magnitude as the 
inverse free propagator, i.e., $\d{\Pi}\sim (gT)^2$, while
the effective self-energy is of an order of magnitude higher, i.e., 
$\s{\Pi}\sim g(gT)^2$. Since the momentum running inside $\s{\Pi}$ is soft,
to calculate it, we have to use effective vertices and propagators
instead of their bare (tree) counterparts.
This ensures the correct expression for the $g(gT)^2-$correction
to the inverse gluon propagator and, in particular, that this correction
is gauge-independent [4].

The hard-thermal-loop $\delta\Pi$ is already known from the literature [2,4].
It is real and contributes to the determination of the spectrum of the
soft gluonic excitations to leading order $gT$. More explicitly,
$\d{\Pi}$ is gauge-invariant and can be expressed in terms of only two 
independent scalar functions, i.e.,:
$$
 \eqalign{\d\Pi^{00}(P)& = \d\Pi_{l}(P)\ ;\qquad 
 \d\Pi^{0i}(P) = - {p_0 p^i\over p^2}\ \d\Pi_{l}(P)\ ;\cr
 \d\Pi^{ij}(P) &= (\delta^{ij}-\h{p}^i \h{p}^j)\ \d{\Pi}_{t}(P)
 + \h{p}^i \h{p}^j\ {(p_0)^2\over p^2}\ \d\Pi_{l}(P)\ .\cr}\eqno(4)
$$
The expressions of $\delta\Pi_{l}(P)$ and 
$\delta\Pi_{t}(P)$ read [2]:
$$
 \d\Pi_{l}(P)=3 m_{g}^2\ Q_{1}\Big({ip_0 \over p}\Big)\ ;\qquad
 \d{\Pi}_{t}(P)= {3\over 5} m_{g}^2\ \Big[Q_{3}\Big({ip_0 \over p}\Big)
 - Q_{1}\Big({ip_0 \over p}\Big) - {5\over 3}\Big]\ ,
 \eqno(5)
$$
where the $Q_n $ are Legendre functions of the second kind. The thermal
gluonic mass is given explicitly by $m_g=\sqrt{N_c+(1/2)N_f}\,gT/3$, where
$N_f$ is the number of flavors.

The effective propagator for soft gluons that intervenes in the calculation
of the effective self-energy is obtained by inverting (2) while disregarding
$\s\Pi$. In the strict Coulomb gauge, its nonzero components are
$\s{\Delta}_{C}^{00}(P)= \s{\Delta}_{l}(P)$ and $\s{\Delta}_{C}^{ij}(P)
= (\delta^{ij}-\h{p}^i\h{p}^j) \hskip2pt\s{\Delta}_{t}(P)$, where
$\s{\Delta}_{t}$ and $\s{\Delta}_{l}$ are given by:
$$
 \s{\Delta}_{t}(P) = {1\over{P^2 - \d{\Pi}_{t}(P)}}\ ; \qquad 
 \s{\Delta}_{l}(P) = {1\over{p^2 - \d{\Pi}_{l}(P)}}\ .    \eqno(6)
$$
After analytic continuation to real energies, the pole in $\omega$ 
of $\s\Delta_{t(l)}$ yields the dispersion relation $\omega_{t(l)}(p)$ 
of the transverse (longitudinal) gluons to order $gT$. The 
longitudinal dispersion equation can be rewritten in the following
explicit manner:
$$
 \ln{\omega_l+p\over \omega_l-p}={2p\over 3\omega_l}\
 \Big(3+{p^2\over m_g^2}\Big)\ .\eqno(7)
$$
One finds for soft longitudinal gluons:
$$
 \omega_l(p)= m_g\,\Big[1+ {3\over 10}\,\Big({p\over m_g}\Big)^2
 - {3\over 280}\,\Big({p\over m_g}\Big)^4 + {1\over 6000}\,
 \Big({p\over m_g}\Big)^6
 +{489\over 43120000}\,\Big({p\over m_g}\Big)^8 + \dots\Big]\ . \eqno(8)
$$

To order $gT$, the gluons are not damped. The leading order of the damping
rates is obtained after is included in the dispersion equations the
contribution from the effective self-energy $^{*}\Pi$, which of course has
a more complicated structure than that of $\delta\Pi$. One shows that in
the strict Coulomb gauge, it also depends on two scalar
functions such that the mass--shell condition that determines the
longitudinal dispersion relation is:
$$
 p^2 - \d\Pi_{l}(-i\Omega_{l},p) - \s\Pi_{l}(-i\Omega_{l},p) 
 = 0\ ,\eqno(9)
$$
where $\s\Pi_l=\s\Pi^{00}$. The longitudinal damping rate is defined
by $\gamma_l(p) \equiv -{\rm Im}\ \Omega_l(p)$. Since it is $g$-times
smaller than the energy $\omega_l(p)$, we can write from (9):
$$
 \gamma_{l}(p) =  
 {{{\rm Im} \s\Pi_{l}(-i\omega, p)}\over{{\partial\over{\partial\omega}}
 \d\Pi_l(-i\omega, p)}}\Big{|}_{\omega=\omega_l(p)+i0^{+}} \ .   \eqno(10)
$$
The denominator in (9) is easy to get from (5). One has
$\partial/\partial\omega\,\d\Pi(-i\omega,p)|_{\omega=\omega_l(p)}=
-2m_g(p/m_g)^2+\dots\ $. This means that in order to get $\gamma_l(p=0)$,
we have to expand the imaginary part of the longitudinal self-energy to
order $p^2$.

In the Coulomb gauge, the only diagrams that contribute to
${\rm Im}\s\Pi_l(P)$ above the light cone are the three-gluon
($3g$) and four-gluon ($4g$) one loop-diagrams with soft internal
momentum [4]. Hence we write:
$$
 \eqalignno{{\rm Im}\ \s\Pi_l(P)= &
 -{g^{2}N_{c}\over 2}\ {\rm Im\ Tr_{soft}}\
 \big[\s\Gamma^{00\lambda\sigma}(P,-P,K,-K)
 \ \s\Delta_{C}^{\lambda\sigma}(K)\cr & +
 \s\Gamma^{\sigma 0\lambda}(-Q,P,-K)
 \ \s\Delta_{C}^{\lambda\lambda'}(K) 
 \ \s\Gamma^{\lambda' 0\sigma'}(-K,P,-Q) 
 \ \s\Delta_{C}^{\sigma'\sigma}(Q)\big]\ ,   &(11)\cr}
$$
where $K$ is the internal loop-momentum, $Q=P-K$ and Tr $\equiv
T\sum_{k_0}\int {d^{3}k\over (2\pi)^3}\ $. The subscript `soft'
means that only soft values of $k$ are allowed in the integral.
Eq (11) is what one would normally write for the $3g$ and $4g$ contributions
to the imaginary part of the gluon self-energy, except that everywhere,
tree quantities are replaced by the corresponding effective ones.
The gluonic effective vertices can be written in the following form:
$$
 \s\Gamma^{(n)} = \Gamma^{(n)} +\ \d\Gamma^{(n)}\ ; \qquad
 n=3,4\, \ , \eqno(12)
$$
where the first term is the tree QCD gluon vertex and the second one sums up
the contributions from hard thermal loops with $n$ external legs.
In the case $n=3$ it can be written as:
$$
 \d\Gamma^{\mu\nu\lambda}(-Q,P,-K) = -\d\Gamma^{\mu\nu\lambda}(-K,P,-Q)=
 3m_{g}^2 \int{d\Omega_S\over 4\pi}
 {{S^\mu S^\nu S^\lambda}\over PS} \Big({iq_0 \over QS} 
 -{ik_0\over KS}\Big)\ ,\eqno(13)
$$
where $S\equiv (i, \h\b s)$ and $\Omega_S$ is the 
solid angle of the unit three-vector $\h\b s$. Also, 
$PS=ip_0+ \b p.\h\b s$, etc. In the case $n=4$ we have:
$$
 \d\Gamma^{\mu\nu\lambda\sigma}(P,-P,K,-K)= 3 m_g^2
 \int{d\Omega_S\over 4\pi} {{S^{\mu} S^{\nu} S^{\lambda} S^{\sigma}}
 \over{PS\ KS}}\Big({ip_0-ik_0\over PS-KS}-{ip_0+ik_0
 \over PS+KS}\Big)\ . \eqno(14)
$$
To be complete, we give the expression of the $3g$ tree vertex:
$$
 \eqalign{\Gamma^{\mu\nu\lambda}(-Q,P,-K)&=
 -\Gamma^{\lambda\nu\mu}(-K,P,-Q)\cr
 &= (P+K)^{\mu}\,\delta^{\nu\lambda}+(Q-K)^{\nu}\,
 \delta^{\lambda\mu}-(P+Q)^{\lambda}\,\delta^{\mu\nu}\ , \cr}\eqno(15)
$$
and that of the $4g$ tree vertex:
$$
 \Gamma^{\mu\nu\lambda\sigma}(P,-P,K,-K)=
 2\delta^{\mu\nu}\,\delta^{\lambda\sigma} - \delta^{\mu\lambda}\,
 \delta^{\nu\sigma} - \delta^{\mu\sigma}\,\delta^{\nu\lambda}\ . \eqno(16)
$$

From eq (11) and the discussion before eq (6), we have:
$$
 \eqalignno{&{\rm Im}\ \s\Pi_l(P)=-{{g^2N_c}\over 2}\
 \r{Im}\ T\sum_{k_0} \int{{d^3k}\over{(2\pi)^3}}\
 \big[\s\Gamma^{0000}(P,-P,K,-K)\ \s\Delta_l(K)\cr
 &+\s\Gamma^{00ij}(P,-P,K,-K)\ (\delta^{ij}-\h{k}^i\h{k}^j)\ \s\Delta_t(K)\cr
 &+\s\Gamma^{000}(-Q,P,-K)\ \s\Delta_l(K)\ \s\Gamma^{000}(-K,P,-Q)\
 \s\Delta_l(Q)\cr
 &+\s\Gamma^{00i}(-Q,P,-K)\ (\delta^{ij}-\h{k}^i\h{k}^j)\ \s\Delta_t(K)\
 \s\Gamma^{j00}(-K,P,-Q)\ \s\Delta_l(Q)\cr
 &+\s\Gamma^{j00}(-Q,P,-K)\ \s\Delta_l(K)\
 \s\Gamma^{00i}(-K,P,-Q)\ (\delta^{ij}-\h{q}^i\h{q}^j)\ \s\Delta_t(Q)\cr
 &+\s\Gamma^{n0i}(-Q,P,-K)\ (\delta^{ij}-\h{k}^i\h{k}^j)\ \s\Delta_t(K)\
 \s\Gamma^{j0m}(-K,P,-Q)\ (\delta^{mn}-\h{q}^m\h{q}^n)\
 \s\Delta_t(Q)\big]\ .&(17)\cr}
$$
There are six contributions: two from the $4g$ vertices
and four from the $3g$ vertices. We will take henceforth
$m_g\equiv 1$ and all momenta and energies are in units of it.
This simplifies considerably the expressions we work with and, if
and when needed, the $m_g$-dependence can easily be recovered in the
final results. Each contribution has to be calculated separately.
As an illustration, let us discuss how we manipulate the contribution
from the $3g$ vertices in which the two effective propagators involved
are both longitudinal. The corresponding term in (17) is:
$$
 \r{Im}\ \s\Pi_{l3gll}(P)=-{g^2N_c\over 2}\ \r{Im}\ T\sum_{k_0}
 \int{d^3k\over(2\pi)^3}\ \s\Gamma^{000}(-Q,P,-K)\
 \s\Delta_l(K)\ \s\Gamma^{000}(-K,P,-Q)\ \s\Delta_l(Q)\ ,\eqno(18)
$$
in a clear notation. Using eqs (12), (13) and (15), we have:
$$
 \eqalign{\r{Im}\ \s\Pi_{l3gll}(P)=
 -\,{9\over 2}\,g^2N_c&\,\r{Im}\,T\sum_{k_0}
 \int{d^3k\over(2\pi)^3}\ \s\Delta_l(K)\ \s\Delta_l(Q)\cr
 &\times\,\Big[\Big(\int{d\Omega_S\over 4\pi}{ik_0\over{PS\,KS}}\Big)^2
 -\int{d\Omega_S\over 4\pi}{ik_0\over PS\,KS}
 \int{d\Omega_S\over 4\pi}{iq_0\over PS\,QS}\Big]\ .\cr}\eqno(19)
$$

We need an expression for the two solid-angle integrals involved
in (19). For this purpose, we use the simple expansion $ 1/PS=
1/ip_0\ (1-\b p .\h\b s/ ip_0-\b p .\h\b s^2/p_0^2+\dots)\,$,
which is valid as long as $p<|ip_0|$, a condition satisfied in the
region $p<m_g$ before analytic continuation to real energies and after.
We then perform the angular integrals straightforwardly and get:
$$
 \eqalign{\int{d\Omega_S\over 4\pi}{ik_0\over PS\,KS}=&
    {ik_0\over ip_0\,k}\,\Big[\,Q_{0k}-{px\over ip_0}\,
    \Big(1-{ik_0\over k}\,Q_{0k}\Big)\cr
    &-{p^2\over p_0^2}\,\Big[-x^2{ik_0\over k}\,
    \Big(1-{ik_0\over k}Q_{0k}\Big)
    +{1\over 2}(1-x^2)\,\Big({ik_0\over k}
    +\Big(1+{k_0^2\over k^2}\Big)\,Q_{0k}\Big)\Big]
    +\dots\Big]\ , \cr}\eqno(20)
$$
where $Q_{0k}$ stands for $Q_0({ik_0\over k})$ and
$x=\cos\,(\h\b p,\h\b k)\,$. We obtain a similar expression for the
other solid-angle integral: we only need to replace $K$ by $Q$.

The next step is to multiply the expressions for the solid-angle
integrals as needed in eq (19) while still expanding in powers
of $p$. Then we integrate over the solid angle of $\h\b k$ and the
odd-powered terms in $p$ cancel out. For the first piece in (19) we get:
$$
 \eqalign{-9\,g^2N_c\,\r{Im}\ T\sum_{k_0}
 &\int{d^3k\over(2\pi)^3}\ \s\Delta_l(K)\ \s\Delta_l(Q)
 \ \Big(\int{d\Omega_S\over 4\pi}{ik_0\over PS\,KS}\Big)^2=
 {9\,g^2N_c\over 2\pi^2}\ \r{Im}\ T\sum_{k_0}
 \int_0^{\infty}{k^2\over -p_0^2}\,dk \cr&
 \times \s\Delta_l(K) \Big[\,{k_0^2\over k^2}\,Q_{0k}^2
 +{p^2\over 3}\,\Big[\,{k_0^2\over k^2}\,Q_{0k}^2\Big({1\over k}\,\partial_k
 +{1\over 2}\,\partial_k^2\Big)+{2\over ip_0}\,{k_0^2\over k^2}\,
 Q_{0k}\Big(1-{ik_0\over k}\,Q_{0k}\Big)\,\partial_k\cr &
 -{1\over p_0^2}\,\Big({k_0^2\over k^2}\,\Big(1-{ik_0\over k}\,Q_{0k}\Big)^2
 -2\,{ik_0^3\over k^3}\,Q_{0k}\Big(1-{ik_0\over k}\,Q_{0k}\Big)\cr&
 +2\,{k_0^2\over k^2}\,Q_{0k}\Big({ik_0\over k}
 +\Big(1+{k_0^2\over k^2}\Big)\,Q_{0k}\Big)\Big)\Big]
 +\dots\Big]\ \s\Delta_l(q_0,k)\ ,\cr}\eqno(21)
$$
where $\partial_k$ stands for $\partial/\partial k$. Similarly, the second
piece in (19) reads:
$$
 \eqalign{+9\,g^2N_c\,\r{Im}&\,T\sum_{k_0}
 \int{d^3k\over(2\pi)^3}\ \s\Delta_l(K)\ \s\Delta_l(Q)
 \int{d\Omega_S\over 4\pi}{ik_0\over PS\,KS}
 \int{d\Omega_S\over 4\pi}{iq_0\over PS\,QS}\cr&
 ={9g^2N_c\over 2\pi^2}\,\r{Im}\,T\sum_{k_0}
 \int{k^2 \over -p_0^2}\,dk\ \s\Delta_l(K)\
 \Big[\,\Big(1+{k^2\over 3}\Big)^2+{p^2\over 3}\,\Big[
 \,\Big(1+{k^2\over 3}\Big)\,\Big(1+{iq_0\over ip_0}\Big)\cr&
 -{2k^2\over 9}\,{ik_0\over ip_0}\,
 -{2\over p_0^2}\Big(1+{k^2\over 3}\Big)^2
 +\,{ik_0\,iq_0\,k^2\over 9\,p_0^2}\,
 +\Big(\,\Big(1+{k^2\over 3}\Big)^2\cr&
 +{k^2\over 3}\,\Big(2-{ik_0\over ip_0}+{iq_0\over ip_0}\Big)\,
 \Big(1+{k^2\over 3}\Big)\,\Big)\,{1\over k}\,\partial_k
 +{1\over 2}\Big(1+{k^2\over 3}\Big)^2\,\partial_k^2\Big]
 +\dots\Big]\ \s\Delta_l(q_0,k)\ .\cr}\eqno(22)
$$
In order to eliminate the appearance of $Q_{0k}^2$ in (21) [1],
extra work is needed. For this purpose, we use the expression
for $\s\Delta_l(K)^{-1}$ one obtains using eqs (6) and (5).
After some algebra and disregarding terms that from the outset
do not yield an imaginary part to the effective self-energy
after analytic continuation, we get:
$$
 \eqalign{-9\,g^2N_c&\,\r{Im}\,T\sum_{k_0}
 \int{d^3k\over (2\pi)^3}\ \s\Delta_l(K)\ \s\Delta_l(Q)
 \ \Big(\int{d\Omega_S\over 4\pi}\,{ik_0\over PS\,KS}\Big)^2\cr&
 ={9\,g^2N_c\over 2\pi^2}\,\r{Im}\,T\sum_{k_0}
 \int{k^2 \over -p_0^2}\,dk\,\Big[-\Big(1+{k^2\over 3}\Big)^2\
 \s\Delta_l(K)+{ik_0\over 3k}\,Q_{0k}\cr&
 +{p^2\over 3}\,\Big[\,{1\over p_0^2}Big(\Big(2+{4k^2\over 3}
 +{2k^4\over 9}-{k_0^2 k^2\over 9}\Big)\ \s\Delta_l(K)
 +{1\over 3}\,\Big(-2+{k_0^2\over k^2}\Big){ik_0\over k}\,Q_{0k}\Big)\cr&
 +\Big(-\Big(1+{k^2\over 3}\Big)\,
 \Big(1+{k^2\over 3}\,\Big(1-2{ik_0\over ip_0}\Big)\Big)
 \ \s\Delta_l(K)+{1\over 3}\,\Big(1-2{ik_0\over ip_0}\Big){ik_0\over k}\,
 Q_{0k}\Big)\,{1\over k}\,\partial_k\cr&
 +{1\over 2}\Big(-\Big(1+{k^2\over 3}\Big)^2\ \s\Delta_l(K)
 +{ik_0\over 3k}\,Q_{0k}\Big)\,\partial_k^2\Big]
 +\dots\Big]\ \s\Delta_l(q_0,k)\ .\cr}\eqno(23)
$$
Remember that all momenta and energies are in units of $m_g$, set to one.
Putting the two pieces (22) and (23) together, we find:
$$
 \eqalign{\r{Im}\,\s\Pi_{l3gll}(P)=&{9\,g^2N_c\over 2\pi^2}\ \r{Im}\,
 T\sum_{k_0}\int{k^2\over-p_0^2}\,dk\,\Big[
 {ik_0\over 3k}\,Q_{0k}+{p^2\over 3}
 \Big[\Big(\Big(1+{k^2\over 3}\Big)\,\Big(1+{iq_0\over ip_0}\Big)
 -{ik_0\,k^2\over 3\,ip_0 }\Big)\ \s\Delta_l(K)\cr&
 +{1\over 3\,p_0^2}\,\Big(-2+{k_0^2\over k^2}\Big){ik_0\over k}\,Q_{0k}
 +k^2\,\Big(1+{k^2\over 3}\Big)\ \s\Delta_l(K)\ \partial_k
 +{1\over 3 k}\,\Big(1-2{ik_0\over ip_0}\Big)
 {ik_0\over k}\,Q_{0k}\ \partial_k\cr&
 +{ik_0\over 6k}\,Q_{0k}\ \partial_k^2\Big]\dots\Big]
 \ \s\Delta_l(q_0,k)\ .\cr}\eqno(24)
$$

The next step is to perform the sum over $k_0$. We do this by using the
spectral representations [5] of the different quantities involved.
In particular, we have for the effective propagators:
$$
 \s\Delta_{t,l}(K)=\int_0^{1\over T}\,d\tau\ e^{ik_0\,\tau}
 \int_{-\infty}^{+\infty} d\omega\ \rho_{t,l}(\omega,k)\
 \big(1+n(\omega)\,\big)\ e^{-\omega\tau}\ ,\eqno(25)
$$
where $n(\omega)=1/\big(\exp(\omega/T)-1\big)$ is the Bose-Einstein
distribution and the spectral densities $\rho_{t,l}(\omega,k)$ have
the following form:
$$
 \rho_{t,l}(\omega,k)={Z_{t,l}(k)\over 2\omega_{t,l}(k)}
 \,\Big[\delta\big(\omega-\omega_{t,l}(k)\big)
 -\delta\big(\omega+\omega_{t,l}(k)\big)\Big]
 +\beta_{t,l}(\omega,k)\ \Theta\big(k^2-\omega^2\big)\ ,\eqno(26)
$$
where $\Theta$ is the step function. The two
residues $Z_{t,l}(k)$ and the two cuts $\beta_{t,l}(\omega,k)$
are given in eqs (3.23) and (3.24) of [1]; see also [5].
The sum over $k_0$ can now be performed straightforwardly. The analytic
continuation to real energies is obtained by the subsequent replacement
$ip_0\to\omega_l(p)+i0^+$ and we extract the imaginary part from our
expressions using the known relation $1/(x+i0^+)=\r{Pr}(1/x)-i\pi\delta(x)$.
We find:
$$
 \eqalign{\r{Im}\ \s\Pi_{l3gll}(P) = &{3g^2\,N_c\,T\over 4\pi}
 \int_0^{+\infty} dk\,\int_{-\infty}^{+\infty}{d\omega_1\over \omega_1}\,
 \int_{-\infty}^{+\infty}{d\omega_2\over \omega_2}\
 \delta(\omega_l-\omega_1-\omega_2)\;\Big[-\omega_l\omega_1 k\,\Theta_1\cr&
 - p^2\,\Big[\,-k^2\Big(3+{2\over 3}k^2\Big)\,\rho_{l1}
 +{\omega_1k\over 3}\,\Big(2+{\omega_1^2\over k^2}\Big)\,\Theta_1
 -2k^3\Big(1+{k^2\over 3}\Big)\,\rho_{l1}\,\partial_k\cr&
 +{\omega_1\over 3}\,\big(1-2\omega_1\big)\,\Theta_1\,
 \partial_k+{\omega_1\,k\over 6}\,\Theta_1\,
 \partial_k^2\Big]\;+\;\dots\Big]\,\rho_{l2}\ .\cr}\eqno(27)
$$
In this equation, $\rho_{li}$ stands for $\rho_l(\omega_i,k),\ i=1,2\,,$ and
$\Theta_1=\Theta(k^2-\omega_1^2)\,$. Also, for $\omega$ soft, we have used
$n(\omega)\simeq T/\omega$. Finally, note that $\omega_l$ depends on $p$ and
in principle eq (27) still needs a further expansion. But this will prove
unnecessary.

The other contributions to $\r{Im}\ \s\Pi_l(P)$ are calculated
in a similar manner. The expressions of the two $4g$-contributions
are quicker to obtain whereas more work is needed for the
other $3g$-contributions, especially the $3gtt$-part. We spare
the reader the details and here we give the final expressions.
The two $4g$-contributions read:
$$
 \eqalign{\r{Im}\ \s\Pi_{l4gl}(P)=&{3\,g^2N_c\,T\over 4\pi}
 \int_0^{\infty}dk\,\int_{-\infty}^{+\infty}{d\omega_1\over \omega_1}\,
 \int_{-\infty}^{+\infty}{d\omega_2\over \omega_2}\
 \delta(\omega_l-\omega_1-\omega_2)\
 \Big[\,\omega_l\omega_1k\,\Theta_1\cr&
 -p^2\,\Big[-\omega_1 k\,\Theta_1+{4\over 3}\omega_1 k^2
 \,\epsilon(\omega_1)\,\delta_1
 -{2\over 3}\omega_1^2 k^2\,\epsilon(\omega_1)\,\partial_{\omega_1^2}
 \,\delta_1\Big]
 \,+\,\dots\Big]\;\rho_{l2}\ ;\cr}\eqno(28)
$$
$$
 \eqalign{\r{Im}\ \s\Pi_{l4gt}(P)=&{3\,g^2N_c\,T\over 4\pi}\int_0^{\infty}dk\,
 \int_{-\infty}^{+\infty}{d\omega_1\over \omega_1}\,
 \int_{-\infty}^{+\infty}{d\omega_2\over \omega_2}\,
 \delta(\omega_l-\omega_1-\omega_2)\,
 \Big[-\omega_l\omega_1 k\,\Big(1-{\omega_1^2\over k^2}\Big)
 \,\Theta_1\cr&
 -p^2\,\Big[{\omega_1\over k^3}\,\Big(k^4-{k^2\over 3}
 +{4\over 3}k^2\omega_1-k^2\omega_1^2\Big)\,\Theta_1
 +{2\over 3}\omega_1^2\,\epsilon(\omega_1)\,\delta_1\Big]
 +\dots\Big]\;\rho_{t2}\ ,\cr}\eqno(29)
$$
Where $\epsilon(\omega)$ is the sign function and $\delta_1\equiv
\delta(\omega_1^2-k^2)$. As to the $3g$-contributions,
we have:
$$
 \eqalign{\r{Im}\ \s\Pi_{l3gtl}(P)&=\;\r{Im}\ \s\Pi_{l3glt}(P)=
 {g^2N_c\,T\over 24\pi}\ \int_0^{\infty}dk\,
 \int_{-\infty}^{+\infty}{d\omega_1\over\omega_1}\,
 \int_{-\infty}^{+\infty}{d\omega_2\over\omega_2}\,
 \delta(1-\omega_1-\omega_2)\cr&
 \times p^2\,\Big[\,9k^2\big(k^2-\omega_1^2\big)^2\,\rho_{t1}\,\rho_{l2}
 -{3\,\omega_1\over k}\,\big(k^2-\omega_1^2\big)\,\Theta_1\,\rho_{l2}
 +{3\omega_1\over 2k^3}\,\big(k^2+2\omega_1-\omega_1^2\big)^2
 \,\Theta_1\,\rho_{t2}\Big]\;+\dots\ ,\cr}\eqno(30)
$$
and we have:
$$
 \eqalign{\r{Im}\ \s\Pi_{l3gtt}(P)=&{3g^2N_c\,T\over 4\pi}
 \int_0^{\infty}dk \int_{-\infty}^{+\infty}{d\omega_1\over \omega_1}
  \int_{-\infty}^{+\infty}{d\omega_2\over \omega_2}
  \delta(\omega_l-\omega_1-\omega_2)\,
 \Big[\omega_l\omega_1k\Big(1-{\omega_1^2\over k^2}\Big)\,\Theta_1\cr&
 -p^2\,\Big[{2\over 9}\big(3k^2(1-2k^2)+(1+6k^2)\omega_1\omega_2
 +8\omega_1^2\omega_2^2\big)\,\rho_{t1}
 +{\omega_1\over 3k^3}\,\big(k^2-2k^4-2k^2\omega_1\cr&
 +(k^2-1)\omega_1^2
 +2\omega_1^3+\omega_1^4\big)\,\Theta_1
 +{2k\over 9}\big(1-2k^2-6k^4-2\omega_1+2(1+4k^2)\omega_1^2
 -2\omega_1^4\big)\,\rho_{t1}\,\partial_k\cr&
 +{\omega_1\over 3}\Big(-1+2\omega_1+{\omega_1\over k^2}
 -{2\omega_1^2\over k^2}\Big)\,\Theta_1\,\partial_k
 +{k^2\over 9}\big(1-2k^2+4(-1+k^2)\omega_1
 +6\omega_1^2-4\omega_1^3\big)\,\rho_{t1}\,\partial_k^2\cr&
 -{\omega_1 k\over 6}\Big(1-{\omega_1^2\over k^2}\Big)
 \,\Theta_1\,\partial_k^2\,\Big]+\dots\Big]\,\rho_{t2}\ .\cr}\eqno(31)
$$

We put now the different contributions together. The terms
independent of $p$ cancel out as they should and, after reorganizing the
terms proportional to $p^2$ in a suitable manner, we obtain, using eq (10)
in the limit $p\to 0$:
$$
 \gamma_l(0)={g^2N_cT\over 24\pi}\
 \Big[a_{l0}^{(1)}+a_{l0}^{(2)}\Big]\ , \eqno(32)
$$
where the pure numbers $a_{l0}^{(1)}$ and $a_{l0}^{(2)}$ have the following
analytic expressions:
$$
 \eqalign{a_{l0}^{(1)}=&\;9\,\int_0^{\infty}dk\
 \int_{-\infty}^{+\infty}{d\omega_1\over \omega_1}\,
 \int_{-\infty}^{+\infty}{d\omega_2\over \omega_2}\,
 \delta(1-\omega_1-\omega_2)\,
 \Big[\,k^4\,\rho_{l1}\,\rho_{l2}\cr&
 -k^2\,\big(k^2-\omega_1^2)^2\,\rho_{t1}\,\rho_{l2}
 +2\big(k^2+\omega_1\omega_2)^2\,\rho_{t1}\,\rho_{t2}
 +{\omega_1\over 6k^3}\,\big(k^2-\omega_1^2\big)^2\,
 \Theta_1\,\rho_{t2}\Big]\ ,\cr}\eqno(33)
$$
and:
$$
 \eqalign{a_{l0}^{(2)}=&\;9\int_0^{\infty}dk\,
 \int_{-\infty}^{+\infty}{d\omega_1\over \omega_1}\,
 \int_{-\infty}^{+\infty}{d\omega_2\over \omega_2}\,
 \delta(1-\omega_1-\omega_2)\,
 \Big[\;{\omega_1\over 3}\big(1-2\omega_1\big)\Theta_1\,\partial_k\rho_{l2}
 +{\omega_1 k\over 6}\,\Theta_1\,\partial_k^2\rho_{l2}\cr&
 +{\omega_1\over 3}\Big(1-2\omega_1-{\omega_1^2\over k^2}
 +2{\omega_1^3\over k^2}\Big)\,\partial_k\Theta_1\,\rho_{t2}
 +{\omega_1k\over 3}\Big(1-{\omega_1^2\over k^2}\Big)\,
 \partial_k\Theta_1\,\partial_k\rho_{t2}+{\omega_1\over 6k}
 \big(k^2-\omega_1^2\big)\,\partial_k^2\Theta_1\,\rho_{t2}\cr&
 +\partial_k\,\Big[-{2k^3\over 3}\Big(1+{k^2\over 3}\Big)\,
 (2-\omega_1)\,\rho_{l1}\rho_{l2}
 +{2k\over 9}\big(k^2(1-3k^2)+(1-4k^2)\omega_1\omega_2
 -\omega_1^2\omega_2^2\big)\,\rho_{t1}\rho_{t2}\cr&
 +{2\omega_1^2\over 3}\Big(2+{\omega_1\over k^2}
 -{\omega_1^2\over k^2}\Big)\,\Theta_1\,\rho_{t2}
 +{k^2\over 9}\big(1-2k^2-4(1-k^2)\omega_1+6\omega_1^2-4\omega_1^3\big)
 \,\rho_{t1}\,\partial_k\rho_{t2}\cr&
 -{\omega_1\over 6}\,\partial_k\,\Big(\Big(k- {\omega_1^2\over k}\Big)
 \,\Theta_1\,\rho_{t2}\Big)\Big]
 +{4\over 3}k^2\omega_1\;\epsilon(\omega_1)\;\delta_1\;\rho_{l2}
 -{2\over 3}k^2\omega_1^2\;\epsilon(\omega_1)\;
 \partial_{\omega_1^2}\delta_1\,\rho_{l2}
 +{2\over 3}\omega_1^2\;\epsilon(\omega_1)\;\delta_1\;\rho_{t2}]\ .\cr}
 \eqno(34)
$$

Things are now well set up to pursue our initial discussion. When introducing
this work, we have said that at zero momentum, the longitudinal-gluon damping
rate is to be equal to the transverse-gluon one, i.e.,
$\gamma_l(0)=\gamma_t(0)$. If one calculates the transverse-gluon
damping rate at zero momentum, one finds $\gamma_t(0)={g^2N_cT\over 24\pi}\,
a_{t0}$, where $a_{t0}=a_{l0}^{(1)}$, the expression of which is given in
(33) above, see [2]. Therefore, one automatically expects $a_{l0}^{(2)}$
to vanish. But this seems not trivial an issue at all.
Indeed, take for example the term
$\omega_1/3\ (1-2\omega_1)\, \Theta_1\,\partial_k\rho_{l2}$ in the analytic expression
of $a_{l0}^{(2)}$ given in (34). It turns out that the contribution of
this term to $a_{l0}^{(2)}$ and hence to $\gamma_{l}(0)$ is infrared
divergent. More precisely, if one cuts the lower bound of the $k$-integral
to a small momentum $\eta\ll 1$, then one straightforwardly works out
the contribution of this term to $a_{l0}^{(2)}$ and finds:
$$
 \eqalign{\int_{\eta}^{\infty}dk\ &\int_{-\infty}^{+\infty}
 {d\omega_1\over \omega_1}\,\int_{-\infty}^{+\infty}
 {d\omega_2\over \omega_2}\;\delta\,(1-\omega_1-\omega_2)
 \;{1\over 3}\omega_1\,\big(1-2\omega_1\big)\,\Theta_1\,\partial_k\rho_{l2}
 =-{1\over 6\eta^2}-{1\over 20\eta}+{1\over 20}\cr&
 -{1\over 2}\int_{1/2}^{+\infty}dk\,{1-2k\over k}\Big[
 \Big(3+k^2+{3(1-k)\over 2k}\,\ln(2k-1)\Big)^2+
 \Big({3\pi(1-k)\over 2k}\Big)^2\Big]^{-1}\ +\r{O}(\eta)\ .\cr}
 \eqno(35)
$$
Note that the integral over $k$ in the right-hand side of the above equation
is finite. Thus one clearly sees that this term is infrared divergent.
Other terms have a similar behavior.

Remember that we decided to carry out this work when we noticed that
terms similar to the ones in (34) do appear in the analytic expression
of $a_{t1}$ in eq (1), terms which are infrared divergent too [1].
At this stage, it becomes clear that the appearance of such terms is linked
to the expansion in powers of $p^2$. Besides, their appearance
in the expression of $\gamma_l(0)$ is puzzling to a certain degree
because as we said, $\gamma_l(0)$ is expected to be divergence-free.
It therefore may be that though such terms, when taken individually,
are infrared divergent, when put together, the infrared
pieces cancel out and we are left with a finite contribution, likely
a vanishing one in the case of $a_{l0}^{(2)}$. But it should be clear that
checking through this point is a delicate matter. In any case, it is beyond
the scope of the present work.

If such a cancellation does occur for both
$a_{l0}^{(2)}$ and $a_{t1}$, does it signal to a more basic pattern related to
some aspects of the infrared behavior of hot QCD or is it a mere
`coincidence'? On the contrary, if the infrared divergences persist, is it
possible to devise a sort of finite-temperature infrared-renormalization
scheme that would absorb them or is it that we will have to `live with', so
to speak? If such a scheme is possible, how consistent would it be with
other calculations?

Still in the case where these divergences persist, we can also naturally
ask whether they are linked in some way to the logarithmic behavior
$\ln(1/g)$ discussed in [3]; in other words, whether they are merely
a different manifestation of the same phenomenon. We think it is too
early to tackle thoroughly this interesting question but nevertheless,
we can put forward the following argument. The logarithmic behavior is
obtained when letting $p$ come from above $m_g$ down into the interior
of the soft region, but keeping it larger then what is expected to be
the magnetic scale, i.e., $g^2T$. Therefore, though arguably we may think
of the infrared divergences in $a_{t1}$ as being another expression
of this logarithmic behavior, this cannot be argued for $a_{l0}^{(2)}$ simply
because here $p$ has to be strictly zero and the logarithmic
behavior cannot be extrapolated down to zero momentum. Besides, Since the
infrared divergences we obtain in the case of $\gamma_l(0)$ are closer
in nature to the ones we have for $\gamma_t(p)$, we find it difficult
at this stage to answer this question positively, especially that the
r\^{o}le of the magnetic effects is still somewhat obscure. In any case,
one should not forget that the quark-gluon plasma is expected to
be stable for zero and very small gluonic momenta, and therefore,
any divergence in the gluonic damping rates, if it persists after
scrutiny, has to be adressed from this standpoint.

It seems to us that all the issues we are raising in this work are worth
investigating, especially that as far as we can tell, they haven't been (at
least fully) adressed in such a way previously.

We would like to thank Asmaa Abada and Olivier P\`ene from Orsay for
interesting and quite useful exchanges that actually helped reshape
parts of our discussion.
\bigskip
\line{\bf References\hfil}
\medskip
\item{[1]} A. Abada, O. Azi, and A. Tadji, {\it `Damping rate for transverse
gluons with finite soft momentum in hot QCD'}, ENSK-ph/97-02,
{\tt hep-ph/9710549}, submitted to JHEP.

\item{[2]} E. Braaten and R.D. Pisarski, Phys.~Rev.~{\bf D}42 (1990) R2156.

\item{[3]} R.D. Pisarski, Phys.~Rev.~{\bf D}47 (1993) 5589. See in the
context of QED and with a different approach J.P. Blaizot and E. Iancu,
{\tt hep-ph/9706397}: Phys.~Rev.~{\bf D}55 (1997) 973; Phys.~Rev.~Lett.~76
(1996) 3080. See also in the context of scalar QED M.H. Thoma and
C.T. Traxler, Phys.~Lett.~{\bf B}378 (1996) 233.

\item{[4]} Reference 2; E. Braaten and R.D. Pisarski, Nucl.~Phys.~{\bf B}337 (1990) 569;
Phys.~Rev.~Lett.~64 (1990) 1338; Nucl.~Phys.~{\bf B}339 (1990) 310;
J. Frenkel and J.C. Taylor, Nucl.~Phys.~{\bf B}334 (1990) 199;
R.D. Pisarski, Phys.~Rev.~Lett.~63 (1989) 1129; Physica {\bf A} 158 (1989)
246; Nucl.~Phys.~{\bf A}498 (1989) 423c.

\item{[5]} R.D. Pisarski, Physica {\bf A} 158 (1989) 146;
Nucl.~Phys.~{\bf B}309 (1988) 476.
\vfil\eject\end